\documentclass[graybox]{svmult}

\usepackage{mathptmx}       
\usepackage{helvet}         
\usepackage{courier}        
\usepackage{type1cm}        
\usepackage{makeidx}         
\usepackage{graphicx}        
\usepackage{multicol}        
\usepackage[bottom]{footmisc}

\makeindex             

\begin{document}

\title*{An architecture for distributed ledger-based M2M auditing for Electric Autonomous Vehicles}
\titlerunning{An architecture for distributed ledger-based M2M auditing for EAVs} 
\author{Dragos Strugar, Rasheed Hussain, Manuel Mazzara, Victor Rivera, Ilya Afanasyev and JooYoung Lee}
\authorrunning{D. Strugar, R. Hussain, M. Mazzara, V. Rivera, I. Afanasyev and J. Lee} 
\institute{Dragos Strugar, Rasheed Hussain, Manuel Mazzara, Victor Rivera, Ilya Afanasyev, JooYoung Lee \at Innopolis University, Universitetskaya str. 1, 420500 Innopolis, Russia, \\ \email{\{d.strugar, r.hussain, m.mazzara, v.rivera,  i.afanasyev, j.lee\}@innopolis.ru}}
%
%
\maketitle

\thispagestyle{empty}
\pagestyle{empty}

\abstract{Electric Autonomous Vehicles (EAVs) promise to be an effective way to solve transportation issues such as accidents, emissions and congestion, and aim at establishing the foundation of Machine-to-Machine (M2M) economy. For this to be possible, the market should be able to offer appropriate charging services without involving humans. The state-of-the-art mechanisms of charging and billing do not meet this requirement, and often impose service fees for value transactions that may also endanger users and their location privacy. This paper aims at filling this gap and envisions a new charging architecture and a billing framework for EAV which would enable M2M transactions via the use of Distributed Ledger Technology (DLT).}

\section{Introduction}
\label{sec:1}

Vehicles, once considered mechanical boxes and used as a primary source of commuting, are now transforming into smart cars. With the massive introduction of software, modern cars are equipped with embedded sensors, actuators and computational capabilities, including communication and storage. This surge in the computation, communication, and automotive technologies gave rise to Intelligent Transportation System (ITS) realized through Vehicular Ad hoc NEtworks (VANET) \cite{6427481}. Simultaneously, Electric Propulsion Vehicle (EPVs or EVs) were introduced to address the drastic increase of global greenhouse gas (GHG) emissions. Despite the numerous regulations that were made to address the ecological problem, technological solutions are also necessary to be used in synergy. If we also consider recent advances in the field of Artificial Intelligence (AI) we have all the instruments to potentially reduce human error in transportation systems by building self-driving vehicles. They are capable of autonomously steering, navigating, vehicle agile maneuvering, making decisions, foreseeing potential accidents and acting better than human drivers in some critical situations \cite{hussain2018, zubov2018}. Having such machines at our disposal could result in numerous benefits, including, but not limited to, fewer accidents, reduced traffic congestions, and enablement for so-called Mobility-as-a-Service (MaaS) \cite{jittrapirom2017mobility}. Let's consider these topics in details.

\textbf{Accidents}. In the United States, motor vehicle crashes are one of the leading causes of death, as approximately two million collisions happen every year \cite{nhtsa2018reasons,nhtsa2018leadingcause}. A system which would eliminate the human error in the vehicle collisions would result in a smaller number of vehicle crashes. 

\textbf{Traffic congestion}. It is the cause of numerous negative effects including delays, air pollution and increased chance of collisions. This problem has not been solved yet and is one of the main concerns for big cities and megalopolises. Research has shown that only 5 \% of Autonomous Vehicles on our roads can significantly reduce the stop-and-go traffic waves that can arise due to lane changing \cite{DBLP:journals/corr/SternCMBBCHHPWP17}. 

\textbf{MaaS}.
Mobility-as-a-Service is a concept that delivers users' transportation needs through a service provider (SP) \cite{jittrapirom2017mobility}. The core idea is to have a fleet of Self-driving vehicles (SDVs) which can perform transportation services for their users. This could lead to the decline in car ownership and private car usage and the expansion of the Sharing economy. Companies like Volkswagen are very interested in this type of transportation systems, as it has the potential to change the automobile industry \cite{youtube2017digitalization}.
With the increasing popularity of EV, the need of having a sufficient charging infrastructure appears. Currently, the most widespread option is the use of Charging Stations (CSs) to deliver electrical energy from the power source to the EV, ensuring the appropriate and safe flow of electricity. CSs are the main interface between the vehicle, the utility and the user \cite{cities2012plug}. This means that the service providers have enabled users to charge their vehicle by plugging it into the CS, and then paying for the services provided through the use of credit/debit cards. Although this may seem convenient for human use, they are not suitable for the Machine-to-Machine (M2M) economy \cite{bojkovic2014machine}. The vehicles of the future will be driverless, requiring the absence of human interaction in the charging process (this is a typical scenario for MaaS). In addition, such an approach leads to fees imposed often by trusted third-party authorities. Finally, they violate the bidirectional auditability and are not privacy-aware \cite{DBLP:journals/corr/HussainKNSTO15}. A new system which would solve these issues is desirable. 

\textbf{DLTs and the Tangle}.
A new decentralized architecture and distributed computing paradigm called the Blockchain has gained significant popularity. It represents an immutable ledger system implemented in a distributed way which eliminates the need for a trusted third-party \cite{nakamoto2008bitcoin}. By using technologies relying on distributed ledgers we could increase the transparency of the whole system while maintaining the high level of its security. The research of DLT-related use cases has been actively conducted for multi-agent systems \cite{danilov2018,kapitonov2017}, and one particular area which could be disrupted by the use of this innovative technology is Internet-of-Things (IoT). The main reasons why IoT could benefit from the DLT are M2M value transactions, increased security and the automation of processes. However, technical limitations of the current blockchain architecture do not allow it to be broadly used in this industry, primarily due to high transaction fees and scalability issues \cite{medium2017introductiontoiota}. Machines will need to make micropayments between themselves, but in order to make them feasible, transaction fees should be very minimal, preferably non-existent. In addition, transaction confirmation times should be very short for the network to be scalable and ready to accept new transactions. Recently, the Tangle has been introduced \cite{popov2016tangle}. It is the next step in the evolution of Distributed Ledger Technologies and is an underlying technology of the IOTA cryptocurrency. Based on a concept of a Directed Acyclic Graph (DAG), Tangle still serves as a distributed database, but it has three main advantages: it is scalable, involves no fees and allows for offline transactions. These characteristics of the Tangle and its cryptocurrency make it possible to become the backbone of the emerging Machine Economy. 

With these concepts in mind, the cryptocurrency which uses the Tangle, IOTA, aims to become the backbone of the emerging Machine Economy, by allowing machines to trade resources (electricity in our case) and services with each other without the involvement of the intermediary.

\textbf{Outline of the paper}.
This paper is focusing on the problem of charging Electric Autonomous Vehicles without human mediation using Distributed Ledger Technologies. The problem is open and its solution may have significant impact in shaping the society of the future. In this Section we have presented the general scenario of EAV and Distributed Ledger Technologies, with particular attention to IOTA, the cryptocurrency we are proposing for the payment method. In  Section~\ref{sec:architecture} we present the overall Architecture with discussion of the technological stack while in Section~\ref{sec:usecase} we show the paradigmatic use case and what features the platform should exhibit. Section~\ref{challenges} introduces the technological complexities of the project, Section~\ref{threatmodel} discusses how to protect EAVs and CSs from malicious cheating in serving and billing, and Section~\ref{sec:conclusions} brings out some reflections and plans for the future. In this paper we define the vision and the architecture of our framework. The formal design of automotive system that can be applied to this case, has been described in \cite{Gmehlich13}.  




\begin{figure}[b]
\sidecaption
\includegraphics[width=11.6cm]{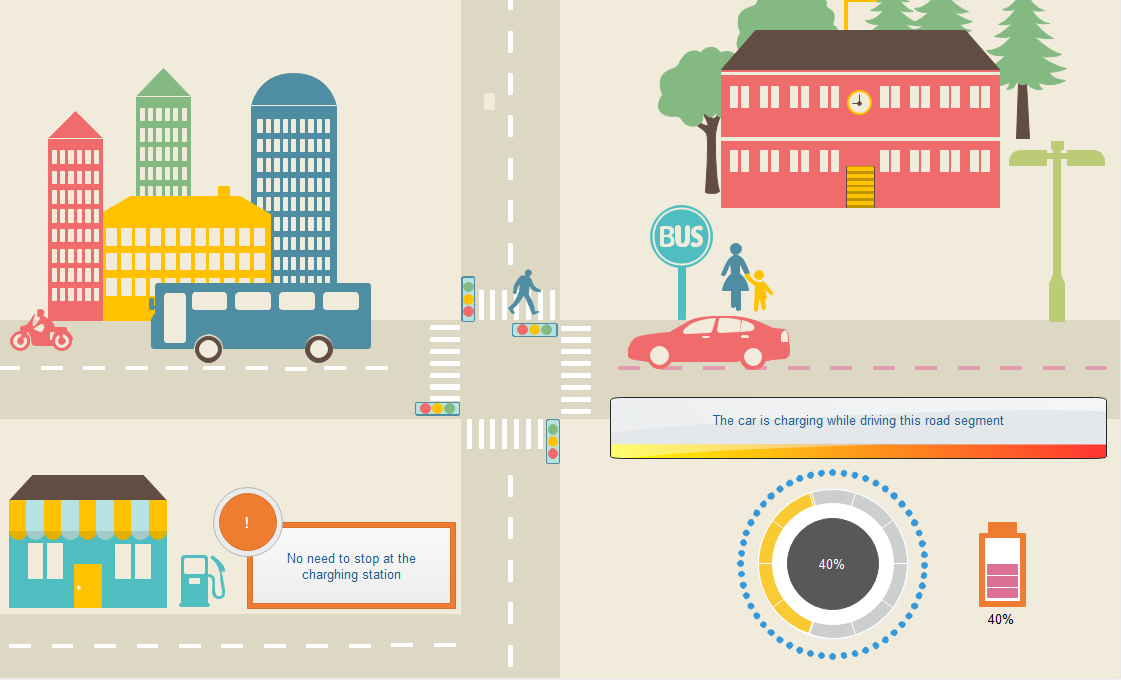}
\caption{Vision of unmanned EAV charging}
\label{fig:scenario}
\end{figure}

\section{Platform Architecture}
\label{sec:architecture}
\begin{figure}[b]
\sidecaption
\includegraphics[width=11.7cm]{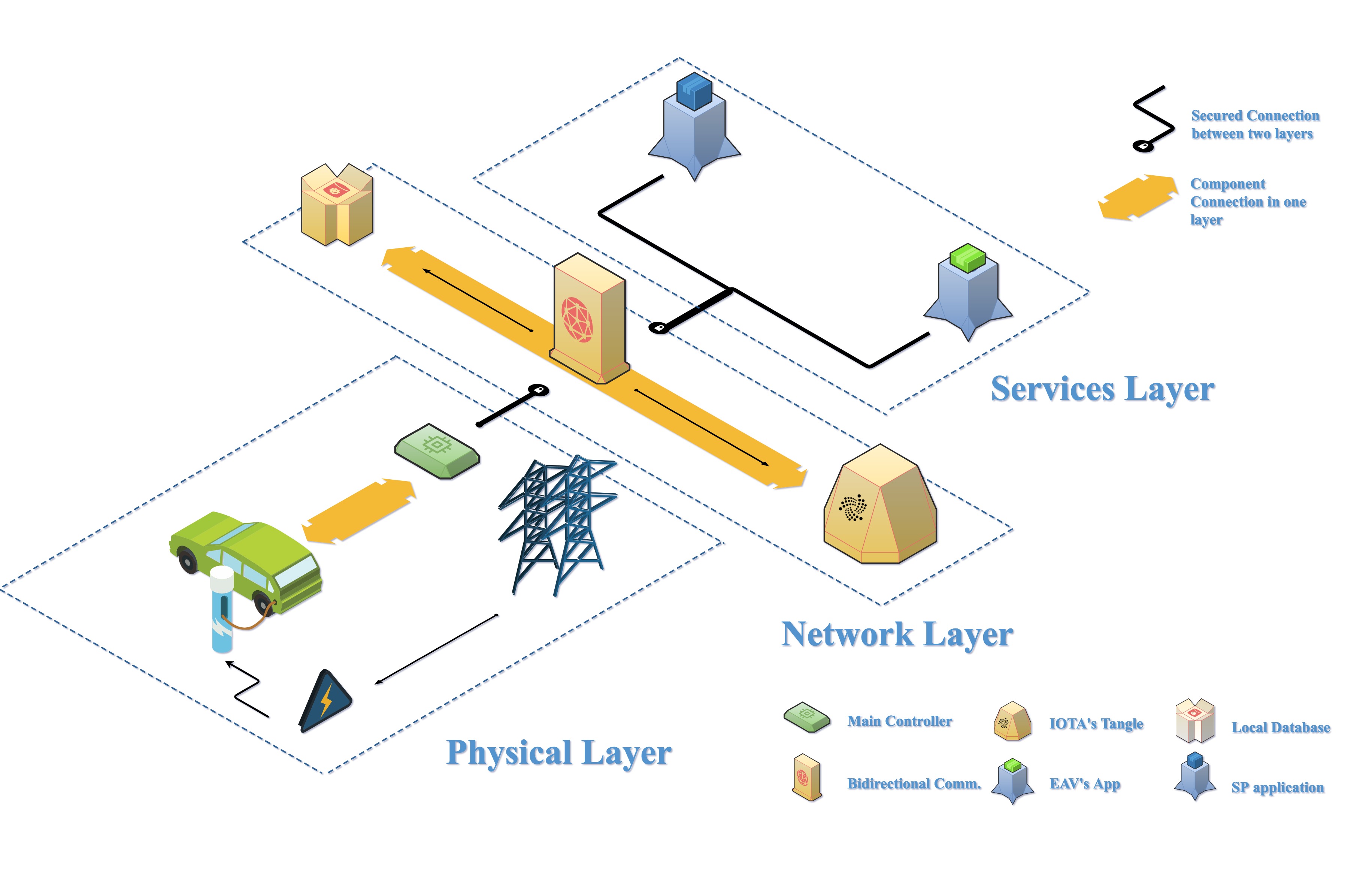}
\caption{System Architecture}
\label{fig:architecture}
\end{figure}

In this section we describe the overall system architecture and its organization as well as the proposed technologies. Figure~\ref{fig:architecture} depicts the architecture consisting of three major layers:
\begin{enumerate}
  \item Physical
  \item Network
  \item Services
\end{enumerate}

\subsection{Physical Layer}
This layer encapsulates all the hardware components embedded in the CS used for sensing and gathering information about the charging process, as well as communicating with external entities. The core parts of this layer are:
\begin{itemize}
  \item kWh meter
  \item Main Controller (MC)
  \item EVSE Controller (EVSEC)
\end{itemize}

The main sensor in the CS is the kWh meter as it measures the amount of electricity consumed by the EAV. Along with the current electricity market price, this data would be constantly used to calculate the total sum that the vehicle would pay to the service provider. The benefit of such an approach is that users get the exact amount of electricity they have paid for.

Moreover, knowing the current electricity rate could help us in adjusting the maximum power the EAV gets charged with, especially in the case of multiple CS using the same power source. The device used to perform such an operation is the Smart Electric Vehicle Supply Equipment (EVSE) \cite{smartevse2014cs}, previously referred to as EVSEC. This could deliver electricity in a controlled (smart) way, and contribute to the development of the Smart Grid \cite{SIANO2014461,medium2017flash}. 

Lastly, we use MC to transfer the information gathered by components of the Physical layer to higher levels of the architecture through the use of a communication protocol specifically designed for IoT, Message Queuing Telemetry Transport (MQTT).

\subsection{Network Layer}
Since our scenario is composed of autonomous, dynamic decision-making CSs and vehicles, we suggest the communication between the entities in the system to be entirely peer-to-peer (P2P). Therefore, peers (EAV and CS) should have equal privileges \cite{yuan2016towards}. In order to allow for P2P communication, we consider DLTs as a possible solution. In particular, we choose IOTA's Tangle as decentralized database that stores both data and value transactions. Communication module built in such a way would benefit from the wide range of possibilities for networking methods, as parties could transfer data and value either offline or online, depending on the protocol used. Since our architecture involves constant M2M communication with subscription behavior, we consider MQTT connectivity protocol to be the most appropriate choice for inter-device communication \cite{hunkeler2008}. In our case, MQTT would ensure rapid transfer of both data and value with high throughput.

In order to allow value transactions in such a system, a more secure solution is needed. Recently, a bi-directional payment channel has been introduced which meets the requirements stated above and conforms to the MQTT protocol by nature. It is called the Flash Channel (FC) \cite{medium2017flashch}, and it allows both resource and service providers to get paid for their services \textit{while} they are performing them. FCs enable real-time streaming of transactions, and allow machines to branch-off the main network, transmit data and value via whichever protocol they find suitable (MQTT in our case), and then merge the results with the main IOTA Tangle. This means that there is no need to wait for transactions to be approved by the IOTA network. By using it, only the starting and ending transactions of the Flash channel will be published to the main network. 

Once machines have finished transacting in FC, the final transaction, denoted as $\mathit{t_f}$, becomes ready to be published on the Tangle. When this happens, $\mathit{t_f}$ has to verify two not yet approved transactions. It does so by performing computational work which is used to verify if the transaction is trusted or not. This process is called the Proof-of-Work (PoW). Similarly, other transactions $\mathit{t_{k \neq f}}$, $\mathit{t_{l \neq k \neq f}}$ will find and verify $\mathit{t_f}$. In order for any transaction $\mathit{t_{x \neq f}}$ to be approved in the Tangle network, two incoming transactions $\mathit{t_y}$ and $\mathit{t_z}$ have to verify it using their PoW. Such an approach removes the need for miners, and therefore enables feeless transactions. In addition, the network benefits from new transactions as they will verify two unapproved ones and thus improve the scalability of the system \cite{popov2016tangle}.

Through the use of the Tangle, FCs and MQTT protocol for message transfer, the Network Layer would be able to transfer sensor data from the Physical Layer to the Services Layer rapidly and securely. 

\subsection{Services Layer}
On the top of the architecture sits the Services Layer which makes use of the two layers described above to deliver services to both consumers and service providers:
\begin{itemize}
  \item Charging services for EAVs
  \item Data Insights for Service Providers
\end{itemize}

As end-consumers of charging services are not humans, but rather EAVs, we envision the application used for payment of the services provided, to be embedded inside the vehicle itself. This application would make the use of Global Positioning System (GPS) and Artificial Intelligence (AI) to locate the most appropriate Charging Stations. In addition, it would be responsible for initiating the charging process, broadcasting the amount of electricity needed and paying for services provided using its IOTA wallet. Lastly, once the charging process is completed, the application would close the payment channel and record the new wallet balance on the main Tangle network. This way, EAV would have its own wallet, and truly embrace the M2M Economy by automating the whole process.

On the other hand, service providers would benefit from the fact that they can analyze the data collected in real-time and gain useful insights. As state-of-the-art NoSQL databases enable low latency, the data collected from the Physical and Network Layers in the process of off-tangle Flash transactions could be used for Machine Learning (ML) scenarios such as graphical demos, predictions and live billing analysis. 
\section{Unmanned EAV charging}
\label{sec:usecase}

Our vision demands investment on the creation of Electric Roads (ER), meaning roads with electric strips that would allow charging on-the-go \cite{guardian2018}. Wireless power transfer (WPT) has been around since Nikola Tesla \cite{Tesla1891}. In such a scenario, EAV would be able to charge without stopping at the station and without the need of any human interaction. The payment framework that we envision is based on IOTA. The ideal scenario is pictorially represented in Figure~\ref{fig:scenario}.
While the wireless method provides numerous benefits, and we consider it to be a more suitable long-term solution for EAV charging, its mass-scale deployment is not yet feasible at the moment \cite{williams2017}. Alternatively, regular Charging Stations (CSs) should be considered.

When the vehicles' Battery Management System (BMS)\footnote{BMS is a combination of sensors, controller, computation and communication hardware with software algorithms designed to estimate the battery percentage of the EV \cite{5609223}.} indicates that the vehicle needs charging, the application embedded in the EAV starts searching for the most suitable CS nearby. It does so by performing AI algorithms based on both Cloud data gathered from various charging points and the current vehicle's location determined by the GPS.

Once arrived at the Charging Point, EAV authenticates with the CS by fetching its wallet address and establishing a secure payment channel, i.e. FC. The address fetched by the vehicle will be used for value transfer once FC has done all the preparatory steps. When the authentication step is completed and the FC is established, both parties have to deposit equal amount of IOTA into a multi-signature address controller by both parties. This step is mandatory as it would prevent parties from refusing to continue to sign the transactions. When these deposits are confirmed, FC does not interact with the network until the channel is closed. After both devices have successfully deposited funds inside a FC, EAV broadcasts the amount of electricity it needs to the CS. Networking layer processes the request and sends the data to the MC of the Physical Layer. When the CS gets this information, the charging process starts. 

While EAV is getting the electricity from the CS, MC is using data from kWh meter to record the amount of electricity consumed. This data is then sent to the EAV's application via MQTT protocol, and the consumer pays the calculated sum through the FC. The transactions are stored in a local (preferably NoSQL) database with low latency. This enables service providers to perform ML algorithms on this data and gain useful insights in real-time. In addition, service providers could also benefit from live-billing analysis and charging process monitoring.

When EAV gets the amount of electricity it has asked for, or when the charging process is manually stopped, FC takes the latest state and spreads the remaining tokens across transacting parties. After each party signs the final bundle, it is then attached to the main network, i.e. the Tangle.

EAV's application records the end of the charging process, and sets its state to inactive.   

\section{Research Challenges}
\label{challenges}

The capabilities of IOTA are unprecedented and it is poised to revolutionize the current micro-economy; however, in the domain of connected and autonomous cars industry, there are challenges that need to be addressed. In this section, we focus on the general IOTA-related and EAV-specific challenges.

\textbf{Adaptation}.
IOTA is still in its infancy and at the moment, it is struggling with growing the Tangle. Nevertheless, IOTA undoubtedly provides many benefits as opposed to blockchain, for instance no transaction fees; however, the adaption of IOTA among businesses is still a challenge and needs to address different issues such as best practices of cryptography (the absence of which may lead to security loopholes), conflict resolutions, approval of new members, and so on. It is, at least at the moment, premature to guarantee its adoption among the consumers and the businesses. In other words, it is still speculative what the future will bring for IOTA, but the signs are convincing that IOTA will become popular for micro-transactions provided that the underlying challenges are addressed. 

\textbf{Performance}.
Mining in blockchain is a time-consuming procedure with the growth in the network which renders it unsuitable for micro-transactions. IOTA - the successor of blockchain, on the other hand, performs better when the network grows. However, at the moment IOTA is not as adopted as much as blockchain and it will need time to grow. Therefore, we will have to wait and see how IOTA evolves in the years to come. Meanwhile, IOTA needs other means of increasing the performance as compared to its other competitors in the market. Furthermore, the performance challenge is also indirectly related to the consumer trust to the IOTA which means that unless consumers are satisfied, the network will not grow and subsequently, the performance will not increase.  

\textbf{Security}.
Currently IOTA has a ``Coordinator" that is managed by IOTA Foundation and it protects users from malicious activities such as DoS and other attacks. However, when the IOTA network grows, IOTA Foundation is planning to remove the coordinator and completely rely on the Tangle itself. The current attacks that target IOTA are mainly based on the DAG where malicious nodes with enough computational power can create a sub-tangle which will cause the IOTA network to approve this malicious transaction. To deter this kind of attack, a strong reputation and validation mechanism is essential. Furthermore, in the current state, the coordinator keeps malicious nodes away from trying to launch different kinds of attacks such as the one discussed above and others such as parasitic chain attacks (where malicious transactions keep trying unless they are approved by the network even though with a very low probability) and other splitting attacks. However, when the network grows and the coordinator is removed from the IOTA network, it is imperative to address these security issues beforehand. 

\textbf{Trust Management}.
Reputation is based on the number of transactions through a particular node in Tangle. The greed for higher score will let malicious nodes to try as many transactions as they can. As aforementioned, this phenomenon could lead to the approval of malicious transactions. Therefore efficient and effective authorization and authentication mechanisms aided with reputation management are essential for IOTA. 

\textbf{Trust in Coordinator}.
The holistic centralized trust in IOTA Foundation where they guarantee the security and validity of transaction could also endanger the efficacy of IOTA. This phenomenon is also related to technical issues with the underlying "nuts and bolts" of the IOTA system, i.e. cryptographic protocols. It is always recommended to use the existing best practices cryptographic protocols; however, IOTA uses its own home-grown hash function curl. It is worth noting that there has been found vulnerabilities in the hash function used by IOTA which may endanger the trust in IOTA network. Furthermore, another problem with IOTA coordinator is that it is a closed source which gives no room for the application developers to maneuver. In order for IOTA network to excel, it has to consider these challenges before they decide to remove the coordinator from the network. 
\section{Threat Model}
\label{threatmodel}
This section discusses how the proposed system would protect EAVs and CSs from malicious cheating during the settlement of payments. One key characteristic that makes the system we proposed in this article different is that it effectively solves the Guest-Host problem (GHP) 
This problem presents the case of a guest who wants to use a hotel room for some number of days and does not want to get billed for the service used it. On the other hand, the host should not get paid less than determined and agreed by both parties. In our case, we consider EAV as the guest, and CS as the host of the GHP problem.

We consider three main scenarios: \textit{(A) Guest tries to get more charge than it has paid for, (B) Host tries to bill the user more than it is agreed,} and \textit{(C) Either side tries to benefit from the interruption of the charging process}.

\subparagraph{\textit{\textbf{A. Guest tries to get more charge than he/she has paid for}}}

\textbf{~~~~~Problem}.
The most common scenario occurs when the guest gets the service as promised, but does not want to pay the settled amount of money. For this very reason, state-of-the-art electric CSs mostly require users to have special service cards tied to their bank accounts and often passports/national IDs\footnote{But it is a prime example of Privacy violation - why service providers are able to see users' detailed private information?}. This way the CS can ensure that the user will get billed. Thus, scenarios where EAVs first get the charge and then pay, are not convenient for service providers and are seldom used.

\textbf{Solution}.
Our approach does not assume that someone should pay first. Instead, we look at the process of charging and billing as token streaming. Key idea is to have a system where \textit{both charging and billing are happening at the same time}. That is, while EAV is getting the charge, it is paying for these services. 
Our solution leverages IOTA, and their bidirectional Flash Channels (FCs) for this specific use case. As we have discussed, FCs enable instantaneous, high-throughput transactions. This fact is crucial in our scenario. Potentially, to be entirely sure that the system is fair, the number of transactions per second can be set.
It would ensure that the users get exact amount of charge for the amount they pay. Additionally, IOTA, by its nature, is privacy-aware, as IOTA transactions do not include any personal information\footnote{IOTA transaction contains: transaction hash, signature message fragment, address of the recipient, value, timestamp, bundle, nonce, and a few other properties that reveal nothing about the identity of transacting parties.te}. 

Thus, the guest (EAV) can not get more charge than it has paid for. And additionally, the whole process of billing has numerous benefits, including, but not limited to: trust, fairness for both parties, transparency and privacy.

\subparagraph{\textit{\textbf{B. Host tries to bill the user more than it is agreed}}} 
\textbf{~~~~~Problem}.
While the state-of-the-art approach does in fact solve one aspect of the GHP (host does get paid), it can be easily seen that this approach is not perfect for the guest. First, EAV exposes its private information, then pays for the service, and in the end, has to trust the CS that it will deliver the promised charging services. And not only that, CS can hide the current market price, and bill users way more than they should be billed. Therefore, users must be able to see the price they are getting the electricity for.

\textbf{Solution}.
Our proposed scheme solves this problem very effectively. At each point in time, users are billed with the amount of money which corresponds to the amount of electricity they get, as there are no fees. The whole process of charging and billing is transparent for the guest of the GHP. If the host tries to bill the guest more than it is agreed, that malicious activity can be easily identified, and the host could have reporting contract violation warning.

\subparagraph{\textit{\textbf{C. Either side tries to benefit from the interruption of the charging process}}}
\textbf{~~~~~Problem}.
One notable case occurs when either the guest or the host decides to quit the charging process by force. In traditional charging station systems, there are very few ways to deal with such a situation. What service providers do, is employ people to regulate the whole process that degrades the M2M concept.

\textbf{Solution}.
As in the previous two cases, payment channels handle also provide a very elegant solution. Either side could terminate the charging process by force, and the only consequence would be that the channel closes. When that happens, CS constructs a bundle that takes the balance of the channel, and divides it amongst the channel’s users accordingly. These bundles are then attached to the network, and neither side benefits from this interruption.
\section{Conclusions}
\label{sec:conclusions}
The emergence of Intelligent Technologies is pervading every moment of our life. While cars themselves may externally look similar to decades ago (apart from design trend consideration), a dramatic revolution is happening inside the cabin. This revolution could be described by the single word: \textit{software}. Software is becoming dominant, at times pervasive, of everyday existence. This certainly demands for high attention on privacy and ethical issues, however expands our possibilities beyond the unthinkable. The same way that we have intelligent portable devices, we have intelligent buildings \cite{Salikhov2016b,Salikhov2016a} that can significantly improve our lives or the lives of people with impaired capacities \cite{Nalin2016}, and so we are aiming at having intelligent cars. In this paper we discussed the synergy of three fundamental technologies: Autonomous Vehicles, Electric Cars and cryptocurrencies. What we propose is their cooperation in order to implement M2M charging with realization of a billing framework for EAVs via IOTA technology. 
 Future work will involve the full realization of the vision here described, at first with a computer graphical simulation, and finally with a full scale development. After these three phases have been completed, we are going to present the project to authorities in order to upgrade the project to a larger scale. Further refinements could include a microservice-based architecture, aimed at increasing scalability \cite{DragoniLLMMS17}, which is already part of our research agenda. At a larger scale, the vision could be expanded foreseeing a marketplace of individuals and companies selling out energy in excess or energy generated precisely for this purpose. Once the technology will be in place, this would create a specific market run on the idea of cryptocurrency as a value-exchanger in a scenario where cars can drive and stop for charge in a completely autonomous manner.

\bibliographystyle{splncs03}
{\bibliography{reference}}

\begin{thebibliography}{10}
\providecommand{\url}[1]{\texttt{#1}}
\providecommand{\urlprefix}{URL }

\bibitem{nhtsa2018reasons}
Critical reasons for crashes investigated in the national motor vehicle crash
  causation survey. Tech. rep., U.S. Department of Transportation, 1200 New
  Jersey Ave, SE, Washington, DC 20590, USA (February 2015),
  \url{https://crashstats.nhtsa.dot.gov/Api/Public/ViewPublication/812115}

\bibitem{nhtsa2018leadingcause}
Traffic safety facts, 2015 - a compilation of motor vehicle crash data from the
  fatality analysis reporting system and the general estimates system. Tech.
  rep., U.S. Department of Transportation, 1200 New Jersey Ave, SE, Washington,
  DC 20590, USA (2015),
  \url{https://crashstats.nhtsa.dot.gov/Api/Public/ViewPublication/812384}

\bibitem{bojkovic2014machine}
Bojkovic, Z., Bakmaz, B., Bakmaz, M.: Machine to machine communication
  architecture as an enabling paradigm of embedded internet evolution. In: Int.
  Con. on ACE. pp. 40--45 (2014)

\bibitem{medium2017flash}
van~den Brink, H.: How elaadnl built a poc charge station running fully on
  iota, and iota only.
  \url{https://medium.com/@harmvandenbrink/how-elaadnl-built-a-poc-charge-station-running-fully-on-iota-and-iota-only-e16ed4c4d4d5}
  (2017)

\bibitem{5609223}
Cheng, K.W.E., Divakar, B.P., Wu, H., Ding, K., Ho, H.F.: Battery-management
  system (bms) and soc development for electrical vehicles. IEEE Trans. Veh.
  Technology  60(1),  76--88 (2011)

\bibitem{cities2012plug}
Cities, C.: Plug-in electric vehicle handbook for public charging station
  hosts. US Department of Energy Publication No. DOE/GO-102012-3275  (2012)

\bibitem{danilov2018}
Danilov, K., Rezin, R., Kolotov, A., Afanasyev, I.: Towards blockchain-based
  robonomics: autonomous agents behavior validation. In: 9th Int. Con. on
  Intelligent Systems. IEEE (2018)

\bibitem{DragoniLLMMS17}
Dragoni, N., Lanese, I., Larsen, S.T., Mazzara, M., Mustafin, R., Safina, L.:
  Microservices: How to make your application scale. In: 1st Convergent
  Cognitive Inf. Tech. pp. 95--104 (2017)

\bibitem{medium2017flashch}
Freiberg, L.: Instant \& feeless \- flash channels.
  \url{https://blog.iota.org/instant-feeless-flash-channels-88572d9a4385}
  (2017)

\bibitem{Gmehlich13}
Gmehlich, R., Grau, K., Iliasov, A., Jackson, M., Loesch, F., Mazzara, M.:
  Towards a formalism-based toolkit for automotive applications. 1st FME
  Workshop on Formal Methods in Software Engineering (FormaliSE)  (2013)

\bibitem{guardian2018}
Guardian: How world's first electrified road charges moving vehicles (2018),
  \url{\url{www.youtube.com/watch?time\_continue\=56I\&v\=VZNHZnyxCm8}}

\bibitem{hunkeler2008}
Hunkeler, U., Truong, H.L., Stanford-Clark, A.: Mqtt-s—a publish/subscribe
  protocol for wireless sensor networks. In: 3rd Int. Con. on Communication
  systems software and middleware and workshops (COMSWARE). pp. 791--798. IEEE
  (2008)

\bibitem{6427481}
Hussain, R., Son, J., Eun, H., Kim, S., Oh, H.: Rethinking vehicular
  communications: Merging vanet with cloud computing. In: 4th Int. Conf.
  CLOUDCOM. pp. 606--609. IEEE (2012)

\bibitem{hussain2018}
Hussain, R., Zeadally, S.: Autonomous cars: Research results, issues and future
  challenges. IEEE Communications Surveys Tutorials pp. 1--1 (2018)

\bibitem{DBLP:journals/corr/HussainKNSTO15}
Hussain, R., Kim, D., Nogueira, M., Son, J., Tokuta, A.O., Oh, H.: {PBF:} {A}
  new privacy-aware billing framework for online electric vehicles with
  bidirectional auditability. CoRR  abs/1504.05276 (2015),
  \url{http://arxiv.org/abs/1504.05276}

\bibitem{jittrapirom2017mobility}
Jittrapirom, P., Caiati, V., Feneri, A., Ebrahimigharehbaghi, S.,
  Alonso-Gonz{\'a}lez, M., Narayan, J.: Mobility as a service. Planning  2(2),
  13--25 (2017)

\bibitem{youtube2017digitalization}
Jungwirth, J.: Digitalization and mobility for all.
  \url{https://youtu.be/x2iBLkfyQuA} (2017)

\bibitem{kapitonov2017}
Kapitonov, A., Lonshakov, S., Krupenkin, A., Berman, I.: Blockchain-based
  protocol of autonomous business activity for multi-agent systems consisting
  of uavs. In: Research, Education and Development of Unmanned Aerial Systems
  (RED-UAS). pp. 84--89. IEEE (2017)

\bibitem{nakamoto2008bitcoin}
Nakamoto, S.: Bitcoin: A peer-to-peer electronic cash system  (2008)

\bibitem{Nalin2016}
Nalin, M., Baroni, I., Mazzara, M.: A holistic infrastructure to support
  elderlies' independent living. Encyclopedia of E-Health and Telemedicine, IGI
  Global  (2016)

\bibitem{popov2016tangle}
Popov, S.: The tangle. cit. on p. 131 (2016)

\bibitem{Salikhov2016b}
Salikhov, D., Khanda, K., Gusmanov, K., Mazzara, M., Mavridis, N.: Jolie good
  buildings: Internet of things for smart building infrastructure supporting
  concurrent apps utilizing distributed microservices. In: 1st Convergent
  Cognitive Inf. Technologies. pp. 48--53 (2016)

\bibitem{Salikhov2016a}
Salikhov, D., Khanda, K., Gusmanov, K., Mazzara, M., Mavridis, N.:
  Microservice-based iot for smart buildings. In: 31st Adv. Inf. Networking \&
  App. Workshops (WAINA) (2017)

\bibitem{medium2017introductiontoiota}
Schiener, D.: Introduction to iota cryptocurrency.
  \url{https://blog.iota.org/a-primer-on-iota-with-presentation-e0a6eb2cc621}
  (2017)

\bibitem{SIANO2014461}
Siano, P.: Demand response and smart grids> a survey. Renewable and Sustainable
  Energy Reviews  30,  461 -- 478 (2014),
  \url{http://www.sciencedirect.com/science/article/pii/S1364032113007211}

\bibitem{smartevse2014cs}
Stegen, M.: Smartevse - smart electric vehicle charging controller.
  \url{https://github.com/SmartEVSE/smartevse} (2014)

\bibitem{DBLP:journals/corr/SternCMBBCHHPWP17}
Stern, R.E., Cui, S., Monache, M.L.D., Bhadani, R., Bunting, M., Churchill, M.,
  Hamilton, N., Haulcy, R., Pohlmann, H., Wu, F., Piccoli, B., Seibold, B.,
  Sprinkle, J., Work, D.B.: Dissipation of stop-and-go waves via control of
  autonomous vehicles: Field experiments. CoRR  abs/1705.01693 (2017),
  \url{http://arxiv.org/abs/1705.01693}

\bibitem{Tesla1891}
Tesla, N.: Experiments with alternate currents of very high frequency and their
  application to methods of artificial illumination. ELECTRICAL REVIEW  (1891)

\bibitem{williams2017}
Williams, J.: Transport innovation of the week: electric charging lanes.
  \url{https://makewealthhistory.org/2017/01/23/transport-innovation-of-the-week-electric-charging-lanes/}
  (2017)

\bibitem{yuan2016towards}
Yuan, Y., Wang, F.Y.: Towards blockchain-based intelligent transportation
  systems. In: IEEE 19th Int. Con. on Intelligent Transportation Systems
  (ITSC). pp. 2663--2668. IEEE (2016)

\bibitem{zubov2018}
Zubov, I., Afanasyev, I., Shimchik, I., Mustafin, R., Gabdullin, A.: Autonomous
  drifting control in 3d car racing simulator. In: 9th IEEE Int. Con. on
  Intelligent Systems. IEEE (2018)

\end{thebibliography}

\end{document}